\documentclass[12pt]{iopart}
\usepackage{iopams}

\usepackage{natbib}
\usepackage{epsfig}
\usepackage{amssymb}

\newcommand{\bm}[1]{\mbox{\ignorespaces\boldmath$#1$}}
\def \r{\bm{r}}
\def \s{\bm{s}}
\def \n{\bm{n}}
\def \k{\bm{k}}
\def \d{{\rm d}}
\def \K{{\;,}}
\def \PP{{\;.}}

\begin{document}

\title{Fluctuation phenomena in crystal plasticity - a continuum model}

\author{Michael Zaiser\footnote[3]{Tel. +44-131-6505671; Fax  +44-131-6513470;
               e-mail M.Zaiser@ed.ac.uk}, Paolo Moretti}
\address{The University of Edinburgh, Institute for Materials and Processes,
The King's Buildings, Sanderson Building, Edinburgh EH9 3JL, UK}

\begin{abstract}
On microscopic and mesoscopic scales, plastic flow of crystals is
characterized by large intrinsic fluctuations. Deformation by
crystallographic slip occurs in a sequence of intermittent bursts
('slip avalanches') with power-law size distribution. In the
spatial domain, these avalanches produce characteristic
deformation patterns in the form of slip lines and slip bands
which exhibit long-range spatial correlations. We propose a
generic continuum model which accounts for randomness in the local
stress-strain relationships as well as for long-range internal
stresses that arise from the ensuing plastic strain
heterogeneities. The model parameters are related to the local
dynamics and interactions of lattice dislocations. The model
explains experimental observations on slip avalanches as well as
the associated slip and surface pattern morphologies.
\end{abstract}


\maketitle

\section{Introduction and Background}

According to the classical paradigm of continuum plasticity,
plastic deformation of crystalline solids resembles a smooth,
quasi laminar flow process. This viewpoint has recently been
challenged by experimental investigations which have demonstrated
that, even in the absence of macroscopic deformation
\begin{figure}[b]
\vspace*{.5cm}
\centerline{\epsfig{file=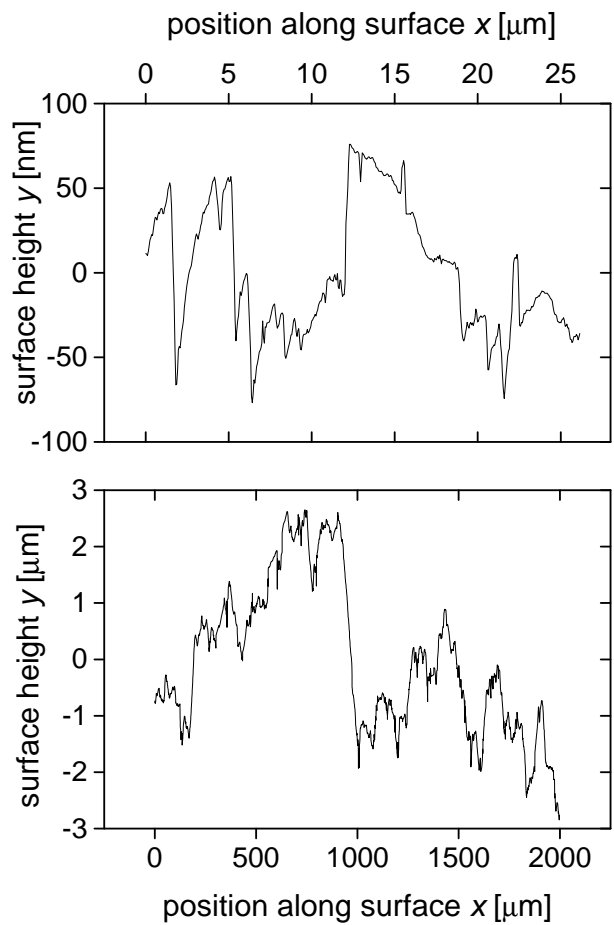,width=8cm,clip=!}}
\caption{Surface profiles of a Cu polycrystal deformed in tension
to a plastic strain $\epsilon = 9.6 \%$. Top: AFM profile; bottom:
SWLI profile. The $x$ direction is parallel to the direction of
the tensile axis. After Zaiser et al. (2004)} \label{profiles}
\end{figure}
instabilities, plastic flow on microscopic and mesoscopic scales
proceeds in a strongly heterogeneous and intermittent manner.
Beyond the microscopic scale, where spatio-temporal localization
of deformation is a trivial consequence of the discreteness of
lattice defects (dislocations), slip localization phenomena extend
over a wide range of mesoscopic scales and may involve the
collective dynamics of very large numbers of defects. This is well
known from the observation of slip lines or slip bands on the
surface of deformed crystals (for an overview, see Neuh\"auser,
1984), where slip steps on the surface manifest the collective
motion of large numbers of dislocations. Recently, it has been
shown that the spatial arrangement of 'slip events' has a fractal
character (Weiss and Marsan, 2003). The consequences of a fractal
distribution of slip on the surface morphology of plastically
deformed crystals have been investigated by Zaiser et al. (2004),
who demonstrated that the surface of plastically deformed Cu
samples develops self-affine roughness over several orders of
magnitude in scale.

\begin{figure}[t]
\centerline{\epsfig{file=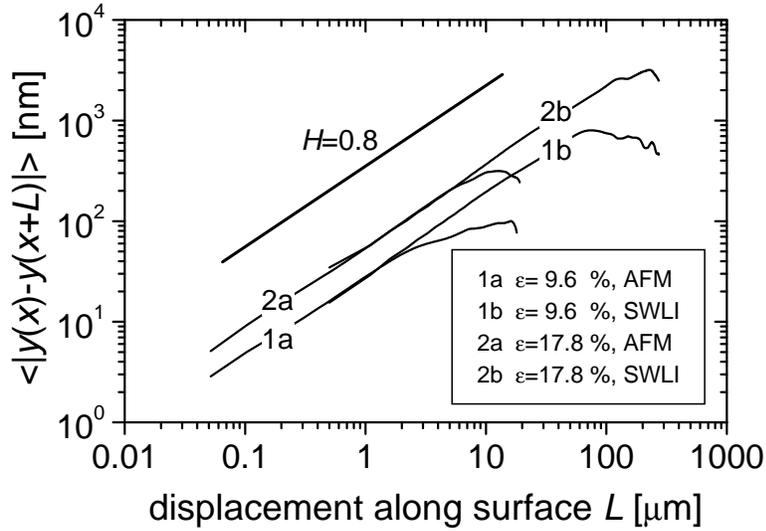,width=10cm,clip=!}}
\caption{Roughness plots (mean height difference vs. distance
along the profile) for AFM and SWLI profiles obtained at strains
of 9.6 and 17.8 \%. The corresponding profiles for $\epsilon = 9.6
\%$ are shown in Figure \ref{profiles}.}\label{roughness}
\end{figure}

Spatio-temporal heterogeneity of slip in spite of macroscopically
stable deformation is illustrated in Figures 1 and 2. Figure
\ref{profiles} shows surface profiles obtained from a Cu
polycrystal with an initially flat surface after deformation to
9.6\% tensile strain. The statistically self-affine nature of the
profiles can be inferred by comparing profiles taken by Atomic
Force Microscopy (AFM, profile length of 25 $\mu$m), and by
Scanning White-Light Interferometry (SWLI, profile length 2 mm).
Over several orders of magnitude, the self-affine behavior can be
quantitatively characterized by a single Hurst exponent $H$ as the
average height difference $\langle |y(x)-y(x+L)| \rangle$ between
two points on a profile increases as a function of their
separation $L$ like $L^H$ with $H \approx 0.8$ (see Figure
\ref{roughness}).

\begin{figure}[t]
\centerline{\epsfig{file=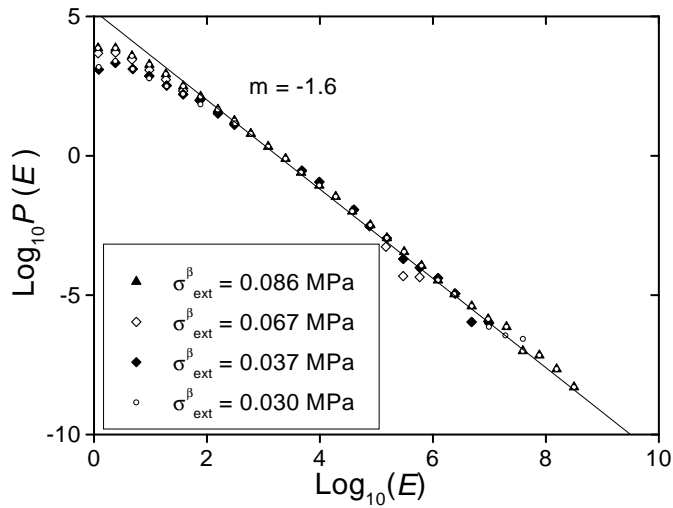,width=10cm,clip=!}}
\caption{Distribution of energy releases in acoustic emission
during creep deformation of ice single crystals; temperature T =
263 K, resolved shear stresses on the basal plane as indicated in
the inset. After Miguel et al. (2001).}\label{aval}
\end{figure}
Information about the temporal dynamics of plastic flow can be
inferred from acoustic emission measurements. Measurements on ice
single crystals reported by Weiss and Grasso (1997) and Miguel et
al. (2001) show temporal intermittency as the acoustic signal
consists of a random sequence of discrete events ('slip
avalanches'). These events exhibit a scale-free size distribution,
as the probability density to observe events with energy release
$E$ decreases according to $p(E) \propto E^{-1.6}$ (Figure
\ref{aval}).

The present paper outlines a simple constitutive model that is
capable of capturing spatial heterogeneity, temporal intermittency
and scale-free behavior in plastic flow of crystals. To this end,
we adopt concepts from the dynamics of random media by assuming
that the deformation of a given volume element occurs through a
random sequence of hardening and softening processes, such that
the local flow stress is a fluctuating function of the local
strain. Physically, these fluctuations result from randomness in
the arrangement of the microstructural defects (dislocations) that
produce the plastic deformation. Because of the crystalline nature
of the medium, plastic deformation takes place by crystallographic
slip, which in the present paper is assumed to occur on a single
slip system only. Fluctuations in the local flow stress lead to
shear strain fluctuations. These, in turn, give rise to long-range
stress redistribution which can be expressed in terms of an
elastic Green's function. Furthermore, dislocation-dislocation
correlations give rise to a local back stress which can be
expressed in terms of a second-order gradient of strain. These
ideas are detailed in Section 2 where a stochastic plasticity
model is presented. In Section 3, the numerical procedure employed
to implement the model is briefly outlined. In Section 4, we
discuss results obtained from the numerical simulations, and
demonstrate that the model correctly accounts for the observed
statistics of 'slip avalanches', the formation and arrangement of
slip lines, and the corresponding deformation-induced surface
morphology.

\section{A Crystal Plasticity Model with Disorder}

\subsection{Basic structure of the model}

We consider the simplest possible model for crystal plasticity
with plastic deformation occurring by crystallographic slip on a
single active slip system. Hence, the plastic distortion tensor is
given by $\bm{\beta}^{\rm p}(\r)=\gamma(\r) \n \otimes \s$ where
$\gamma$ is the shear strain on the slip system, and $\n$ and $\s$
are unit vectors pointing in the direction of the slip plane
normal and the slip direction, respectively. In the following we
assume without loss of generality that the slip direction
corresponds to the $x$ direction of a Cartesian coordinate system
and the slip plane is the $xz$ plane. The driving force for
plastic flow is the shear stress $\sigma_{xy}(\r) =: \tau(\r)$
acting in this slip system. We consider rate-independent
elastic-plastic behavior. Hence, the stress acting in a volume
element located at $\r$ must fulfill the inequality
\begin{equation}
\tau(\r) \le \tau_{\rm f}(\r,\gamma(\r))\PP \label{const}
\end{equation}
Here $\tau_{\rm f}$ is the local flow stress, which depends on
internal state of the volume element under consideration. It is in
general a function of the local strain $\gamma(\r)$ and may also
depend explicitly on the space coordinate $\r$. The stress
$\tau(\r) = \tau_{\rm ext} + \tau_{\rm int}(\r)$ acting from
outside on the considered volume element is a sum of external and
long-range internal stresses. The external stress, which acts as
an external driving force on the system, is assumed space
independent over the region of interest. The space-dependent
internal stress field $\tau_{\rm int}(\r)$ is a functional of the
(in general spatially non-homogeneous) strain field $\gamma(r)$;
it vanishes for a homogeneous deformation state.

If the inequality (\ref{const}) is violated, the local strain
$\gamma(\r)$ increases quasi-instantaneously until Eq.
(\ref{const}) is again satisfied. Impossibility to satisfy this
equation implies ductile failure of the system.

\subsection{Evaluation of long-range internal stresses}

We calculate the internal stresses in an infinite
three-dimensional body with an arbitrary plastic distortion field
$\bm{\beta}^{\rm p}(\r)$. The external stress is assumed to be
zero. (A non-zero external stress simply adds to the internal
stresses.) We start out from the elastic equilibrium equation for
the components $\sigma_{ij}$ of the stress tensor:
\begin{equation}\label{partial_sigma}
-\partial_j\sigma_{ij}=f_i\K
\end{equation}
where $f_i$ are the body forces and sums are performed over
repeated indices. The above equation can be rewritten in terms of
the components $u_i$ of the elastic displacement vector as
\begin{equation}\label{partial_u}
-\partial_jC_{ilkl}\partial_ku_l=f_i,
\end{equation}
where $C_{ijkl}$ are components of Hooke's tensor. The solution of
Equation (\ref{partial_u}) has the form
\begin{equation}
u_i({\bf r})=\int \Gamma_{ik}({\bf r}-{\bf r'})f_k({\bf r'}) d^3r'
\end{equation}
where the Fourier transform of the elastic Green's tensor
$\Gamma_{ik}({\bf r})$ is $\tilde{\Gamma}_{ik}({\bf
k})=[-C_{ijkl}k_jk_l]^{-1}$.

We now first consider the particular problem of a plastically
deformed 'inclusion' where the plastic distortion has a constant
value ${\bf \beta}^{\rm p}$ over a certain volume $V$ and is zero
elsewhere. This {\it inclusion problem} is solved as follows: The
volume $V$ is first cut out of the surrounding matrix and deformed
plastically in order to produce a stress-free strain $\beta^{\rm
p}$. To re-insert it into the matrix, interface tractions are
applied, such that the original shape is restored. According to
Equation (\ref{partial_sigma}) these interface tractions are
\begin{equation}\label{tractions}
f_i=C_{ijkl}\beta^{\rm p}_{kl}\partial_j H_V({\bf r}),
\end{equation}
where the function $H_V({\bf r})$ is equal to unity within $V$ and
zero elsewhere. The volume is then placed in its original position
and relaxed. Elastic relaxation proceeds until the tractions
produced by the relaxation strain $\beta^{\rm r}$ balance those
given by Equation (\ref{tractions}). The Fourier transform of the
corresponding displacement field is
\begin{equation}
u_i^r({\bf k})=i\Gamma_{ik}({\bf k})C_{klmn}\beta^{\rm
p}_{mn}k_lH_V({\bf k})\PP.
\end{equation}
The total elastic distortion is then the sum of the relaxation
strain and the initial distortion $-\beta_{ij}^{\rm p}H_V({\bf
r})$ applied to ``restore'' the original shape before relaxation.
The associated total stress reads $\sigma_{ij}({\bf
r})=C_{ijkl}(\beta^{\rm r}_{kl}-\beta^{\rm p}_{kl}H_V({\bf r}))$
and its Fourier transform is
\begin{equation}
\sigma_{ij}({\bf k})=-C_{ijlm}\left(k_mk_o\tilde{\Gamma}_{ln}({\bf
k}) C_{nopq}+\frac{1}{9}\delta_{lp}\delta_{mq}\right)
\beta_{pq}^{\rm p}\tilde{H}_V\PP
\end{equation}
The angular average
\begin{equation}
\tilde{\Gamma}^0_{ijlm}=\frac{1}{4\pi}\int
C_{ijno}[k_ok_q\tilde{\Gamma}_{np}({\bf
k})C_{pqlm}+\frac{1}{9}\delta_{ln}\delta_{mo}]d\Omega
\end{equation}
does not depend on the modulus $k$ of the wavevector since
$\tilde{\Gamma}_{np}$ scales like $k^{-2}$. The stress hence can
be written as
\begin{equation}\label{tot_stress}
\sigma_{ij}({\bf
k})=[-\tilde{\Gamma}_{ijlm}^0+\tilde{\Gamma}^*({\bf
k}/k)]\beta^{\rm p}_{lm}H_V({\bf k})\K
\end{equation}
where the second term is simply defined by $\tilde{\Gamma}^*({\bf
k}/k)= C_{ijno}[k_ok_q\tilde{\Gamma}_{np}({\bf
k})C_{pqlm}+\frac{1}{9}\delta_{ln}\delta_{mo}]-\tilde{\Gamma}^0_{ijlm}$.

The above procedure can be straightforwardly generalized for an
arbitrary distribution of the plastic distortion ${\bf \beta}^{\rm
p}({\bf r})$ by considering each volume element as a separate
``inclusion''. Equation (\ref{tot_stress}) becomes
\begin{equation}
\sigma_{ij}({\bf
k})=[-\tilde{\Gamma}_{ijlm}^0+\tilde{\Gamma}^*({\bf
k}/k)]\beta^{\rm p}_{lm}({\bf k})
\end{equation}
and, in real space,
\begin{equation}
\sigma_{ij}({\r})=-\tilde{\Gamma}_{ijlm}^0\beta^{\rm p}_{lm}({\bf
r}) +\int\Gamma^*({\bf r}-{\bf r'})\beta^{\rm p}_{lm}({\bf
r'})d^3r'\PP
\end{equation}
The non-local kernel $\Gamma^*({\bf r}-{\bf r'})$ is the inverse
Fourier transform of $\tilde{\Gamma}^*({\bf k}/k)$; it scales like
$1/r^3$ and it can be shown that it has zero angular average. What
has been done so far holds for strain fields that go to zero at
infinite distances. If the asymptotic value of the plastic strain
assumes a non-zero value ${\bf \beta}^{{\rm p},\infty}$, we have
to add the corresponding stress-free strain as follows:
\begin{equation}
\sigma_{ij}({\bf k})=\tilde{\Gamma}_{ijlm}^0[\beta^{{\rm
p},\infty}_{lm}-\beta^{\rm p}_{lm}({\bf r})] +\int\Gamma^*({\bf
r}-{\bf r'})\beta^{\rm p}_{lm}({\bf r'})d^3r'.\label{mf0}
\end{equation}
For a plastic distortion field which has the {\em average} value
$\langle \beta^{\rm p} \rangle$, the asymptotic value $\beta^{{\rm
p},\infty}$ is replaced by the average $\langle \beta^{\rm p}
\rangle$ since the fluctuation contributions average out if
integrated over the infinite contour. Hence, the internal stress
can be envisaged as the sum of a mean-field contribution and a
non-local term with a kernel of zero angular average. For the
purpose of a depinning theory built with this type of non-local
elastic interaction, one may note that in Fourier space the kernel
scales like $k^a$ with $a=0$, i.e., mean-field theory is expected
to be valid in all dimensions.

For later use we give explicit expressions for the case where the
plastic strain is determined by slip on  a single slip system,
$\bm{\beta}^{\rm p} = \gamma(\r) \e_y \otimes \e_x$, and the shear
strain $\gamma$ depends on the $x$ and $y$ coordinates only. (Such
a quasi-two-dimensional model corresponds to a system of straight
parallel edge dislocations, cf. below.) In this case, the internal
shear stress $\tau_{\rm int} = \sigma_{xy}$ is in Fourier space
given by
\begin{equation}\label{stress2dfour}
\langle\tau_{\rm int}(\k)\rangle = -\frac{G}{\pi(1-\nu)}
\gamma(\k)\frac{k_x^2k_y^2}{|\k|^4}\K
\end{equation}
and in real space by
\begin{eqnarray}\label{stress2d}
\langle\tau_{\rm int}(\r)\rangle &=& \frac{G}{2\pi(1-\nu)}\int
\gamma(\r') \left[\frac{1}{(\r-\r')^2} - \frac{8
(x-x')^2(y-y')^2}{(\r-\r')^6}\right]\d^2 \r'\nonumber\\
 &+&
\frac{G}{4(1-\nu)} [\langle\gamma\rangle - \gamma(\r)]\;.
\label{mf}
\end{eqnarray}
Here, $G$ is the shear modulus and $\nu$ the Poisson's ratio. From
these expressions, two points may be noted: (i) The elastic kernel
is not positively definite in real space. (ii) There exist certain
space-dependent strain fluctuations (fluctuations with wavevectors
in the $x$ and $y$ directions) which do not give rise to any
long-range internal stresses. The implications of this will be
discussed below.

\subsection{Dislocation-related stresses and internal-stress fluctuations}

We now discuss the physical origin of the 'flow stress' $\tau_{\rm
f}(\r,\gamma)$. On the microscopic scale, plastic flow of a
crystal is brought about by the motion of linear lattice defects
(dislocations). The flow stress of a small mesoscopic volume (a
volume containing multiple dislocations/dislocation segments)
corresponds to the stress required to move dislocations through
the stress 'landscape' within this volume. In the absence of other
defects, this 'landscape' is created by the dislocations
themselves. The resulting flow stress $\tau_{\rm f} = \delta \tau
+ \tau_{\rm p}$ can be envisaged as a sum of two contributions
[for a more detailed discussion, see Zaiser and Seeger (2002),
Groma, Csikor and Zaiser (2003), Zaiser and Aifantis (2005)]:

(i) A spatially fluctuating stress $\delta \tau(\r)$ which depends
on the positions of the individual dislocations. The average of
the fluctuating stress is zero, and the characteristic 'amplitude'
of the fluctuations is given by
\begin{equation}
\langle \delta \tau(\r)^2 \rangle = K^2 G^2 b^2 \rho(\r)\;,
\end{equation}
where $K^2 = \ln(\xi/b)/[8\pi(1-\nu)^2]$ is a numerical constant
of the order of unity which depends on the elastic properties of
the crystal lattice, and logarithmically on the characteristic
range $\xi \approx 1/\sqrt{\rho}$ of dislocation-dislocation
correlations. The spatial correlation function of the fluctuating
internal stress field is given by
\begin{equation}
\langle \delta \tau(\r) \delta \tau(\r+\r') \rangle = \langle
\delta \tau(\r)^2 \rangle h(\r'/\xi)\;,
\end{equation}
where $h(\r'/\xi)$ is a short-range correlation function with
characteristic range $\xi\approx 1/\sqrt{\rho}$ (Zaiser and
Seeger, 2002).

Dislocation glide increases the local strain and modifies the
fluctuating stresses within the surrounding mesoscopic volume
element. In the absence of detailed information about the
individual dislocation positions, we take this evolution of the
stress 'landscape' into account by envisaging the fluctuating
stress as a random function not only of space but also of {\em
strain}, with the correlation function
\begin{equation}
\langle \delta \tau(\r,\gamma) \delta \tau(\r+\r',\gamma+\gamma')
\rangle = \langle \delta \tau(\r)^2 \rangle
h(\r'/\xi)g(\gamma/\gamma_{\rm corr})\;,
\end{equation}
where $g$ is another short-range function. The 'correlation
strain' $\gamma_{\rm corr}$ can be estimated as the local strain
produced when all dislocations in a volume element move by the
distance $\xi$, $\gamma_{\rm corr} = \rho b \xi \approx b
\sqrt{\rho}$.

(ii) If small groups of dislocations move in a correlated manner,
their mutual interactions tend to homogenize deformation in the
slip plane. This can be described by a 'pile-up stress' $\tau_{\rm
p}(\r)$ which can be approximated by a second-order gradient of
the strain according to
\begin{equation}
\tau_{\rm p}(\r) = \frac{DG}{\rho} \gamma_{xx} \quad(d=2)\;,
\quad\tau_{\rm p}(\r) = \frac{DG}{\rho} [\gamma_{xx} +
\gamma_{zz}]\quad(d=3)\;. \label{gradterm}
\end{equation}
Here, $\rho$ is the total dislocation density and $D$ a constant
of the order of unity [see Groma, Csikor and Zaiser (2003)]. In
Eq. (\ref{gradterm}), $d \in [2,3]$ is the dimensionality of the
model. The first expression ($d=2$) refers to the special case
where deformation is due to the motion of straight edge
dislocations, such that the shear strain $\gamma$ depends on the
$x$ and $y$ coordinates only.

\subsection{Strain hardening}

During deformation, the dislocation density often increases with
increasing strain, leading to an increase of the stress required
to sustain plastic flow (strain hardening). In our model we may
take this into account in two manners:

(a) We explicitly account for the strain evolution of the
dislocation density $\rho$. In the simplest case, this is given by
[see e.g. Zaiser (1998)]
\begin{equation}
\partial_t \rho = \frac{\beta}{b} \sqrt{\rho} \dot{\gamma}\quad,\quad
\rho(\gamma) = \left(\frac{\beta\gamma}{2b}\right)^2\K
\end{equation}
where $\beta \approx 0.02$ is a small non-dimensional parameter.
It follows that the dislocation density is approximately constant
over the 'correlation strain' $\gamma_{\rm corr} \approx
b\sqrt{\rho}$ but increases slowly over larger strain intervals.
This leads to a corresponding increase of the amplitude of the
fluctuating stress $\delta \tau(\r,\gamma)$, which scales in
proportion with the local strain $\gamma(\r)$.

(b) In phenomenological hardening theories, hardening is often
described by a strain dependent 'back stress' which is subtracted
from the locally acting stress. In the simplest case of linear
strain hardening, this stress is given by $\tau_{\rm b} = - \Theta
\gamma$ where $\Theta$ is the hardening coefficient. If we choose
this description of hardening, we keep the amplitude of the
fluctuating stresses constant but subtract a slowly increasing
back stress which is proportional to the local strain.

Actually, both descriptions yield very similar results. As
discussed in the next paragraph, our equations have the structure
of an elastic manifold depinning model, with the fluctuating
stress $\delta \tau$ acting as a random pinning field. Since the
pinning stress in such models is governed by the negative maxima
of the pinning field, a linear increase of the amplitude of the
pinning field with increasing strain, or the subtraction of a
strain-dependent back stress, both have the same consequence: The
stress required to overcome these maxima increases linearly as a
function of the local strain.

\subsection{Plastic flow and elastic manifold depinning}

Collecting all stress contributions discussed in Sections above,
we find that plastic flow occurs ($\gamma$ increases
quasi-instantaneously) as soon as the inequality
\begin{equation} \tau_{\rm ext} + \tau_{\rm int}(\r) +
\frac{DG}{\rho} [\gamma_{xx}+\gamma_{yy}] + \delta \tau(\r,\gamma)
\le 0\label{depin1}
\end{equation}
is violated. The asymptotic behavior of this model for positive
external stresses corresponds to that of the equation of motion
\begin{equation}
\frac{1}{B} \partial_t \gamma(\r) = \tau_{\rm ext} + \tau_{\rm
int}(\r) + \frac{DG}{\rho} [\gamma_{xx}+\gamma_{yy}]  + \delta
\tau(\r,\gamma)\PP\label{depin2}
\end{equation}
in the rate-independent limit $B \to \infty$. Formally, Eq.
(\ref{depin2}) can be envisaged as describing the overdamped
dynamics of an elastic manifold with coordinates $(\gamma,\r)$
which moves in the $\gamma$-direction through a random medium. In
the two-dimensional case where $\gamma=\gamma(x,y)$ depends only
on the $x$ and $y$ coordinates ($d=2$), our model may be
considered a continuum approximation of a quasi-two-dimensional
system of straight parallel dislocations. If the strain depends on
all three spatial coordinates, and the model mimics the
large-scale behavior of a system of three-dimensionally curved
dislocations. In either case, the model can be formally envisaged
as describing the motion of a $d$-dimensional elastic manifold
through a disordered $d+1$-dimensional medium which exerts a
fluctuating pinning force $\delta \tau$. Due to the infinite range
of the interaction kernel governing the internal stress $\tau_{\rm
int}(\r)$, one expects the second-order gradient term in Eqs.
(\ref{depin1},\ref{depin2}) to be irrelevant for the large-scale
behavior, and the model exhibits mean-field behavior irrespective
of the dimensionality of the 'manifold'.

Two differences may be noted with respect to conventional models
of elastic manifold depinning: (i) If we account for hardening,
either the mean value or the amplitude of the pinning field
increases as the manifold advances, i.e., we are dealing with a
net 'uphill' motion of the manifold. The consequences of this will
be discussed in Section 4.2. (ii) As can be directly seen from Eq.
(\ref{mf}), the elastic kernel mediating the long-range
interaction is not positively definite. Hence, the 'no
passing'-theorem (Middleton 1992) need not hold and the existence
of a unique depinning threshold is not a priori guaranteed. In
spite of these caveats, numerical investigation of the model shows
depinning-like behavior, as demonstrated in Section 4.

\section{Numerical Implementation}

For numerical implementation of the model we focus on the
two-dimensional case. (Because of the mean-field nature of the
long-range interactions, a generalization to three dimension is
not expected to make any significant difference). We use a lattice
automaton model where we discretize space in terms of a
two-dimensional array of cells, each cell representing a
mesoscopic volume element whose edge length $l \approx \xi$ is of
the order of the correlation length of the stress fluctuations.
Accordingly, the fluctuating stresses in different volume elements
are assumed as statistically independent random variables. In a
similar spirit, the local strains are discretized in units of the
correlation strain $\gamma_{\rm corr} \approx b\sqrt{\rho}$, and
the fluctuations at different strain steps are considered
statistically independent. A bulk system is mimicked by imposing
periodic boundary conditions in both directions. The total
dislocation density $\rho$ and, hence, the statistical properties
of the fluctuating internal stress field are assumed spatially
homogeneous. A non-dimensional and dislocation density independent
formulation is achieved by scaling all lengths in proportion with
$\xi \approx 1/\sqrt{\rho}$, all stresses in proportion with $K G
b \sqrt{\rho}$, and all strains in proportion with $b/\xi \approx
b\sqrt{\rho}$.

The procedure for implementing the random function
$\delta\tau(\gamma,\r)$ in Eq. (\ref{depin1}) is here discussed in
the absence of hardening. To each cell we assign initially a
negative random value of $\delta\tau$ drawn from the negative half
of a Gaussian distribution with unit variance [Eq. (3)]. This
ensures that there is no plastic flow at zero external stress; in
physical terms, it implies that initially all dislocations are
trapped in configurations where they are held up by negative
(back) stresses. Whenever the strain within a cell increases, we
assign a new (positive or negative) random value of $\delta \tau$
to this cell, which we draw again from a Gaussian distribution of
unit variance. We note in passing that the Gaussian shape of the
probability distribution of $\delta \tau$ is not crucial; other
distributions with finite first and second moments produce similar
results.

In each simulation we increase the applied stress $\tau_{\rm ext}$
from zero in small increments $\Delta \tau_{\rm ext}$. After a
stress increment we check for all volume elements whether Eq.
(\ref{depin1}) is violated. In all volume elements where this
happens, we increase the local strains by a unit amount and assign
new values of $\delta \tau$. After this simultaneous update
procedure is completed , we compute the internal stress field
$\langle \tau_{\rm int}(\r) \rangle$ corresponding to the new
strain pattern by using the periodically continued elastic Green's
function of a two-dimensional isotropic medium, Eq. (\ref{mf}),
and evaluate the gradient-dependent stress contribution using the
discrete second-order gradient. Once the new internal stresses are
computed, we check again for all volume elements whether Eq.
(\ref{depin1}) is now satisfied and, if not, increase the local
strains etc.. The process is repeated until the local stress
everywhere falls below the local yield stress, or until the
average strain within the system exceeds a prescribed maximum
value at which we terminate our simulation. If the system settles
into an equilibrium configuration before reaching the terminal
strain, we increase the external stress further by $\Delta
\tau_{\rm ext}$, and so on. This procedure implements an
adiabatically slow drive, provided that the external stress
increment $\Delta \tau_{\rm ext}$ is chosen small enough (the
practical criterion being simply that the results do not change if
the stress increment is further decreased).

\section{Results and Discussion}

\subsection{Stress-strain characteristics and avalanche dynamics in the absence of hardening}

We first study the model in the absence of hardening. Then, the
'pinning stress' $\delta\tau$ has stationary stochastic
properties.
\begin{figure}[b] \hspace*{-1.5cm}
\centerline{\epsfig{file=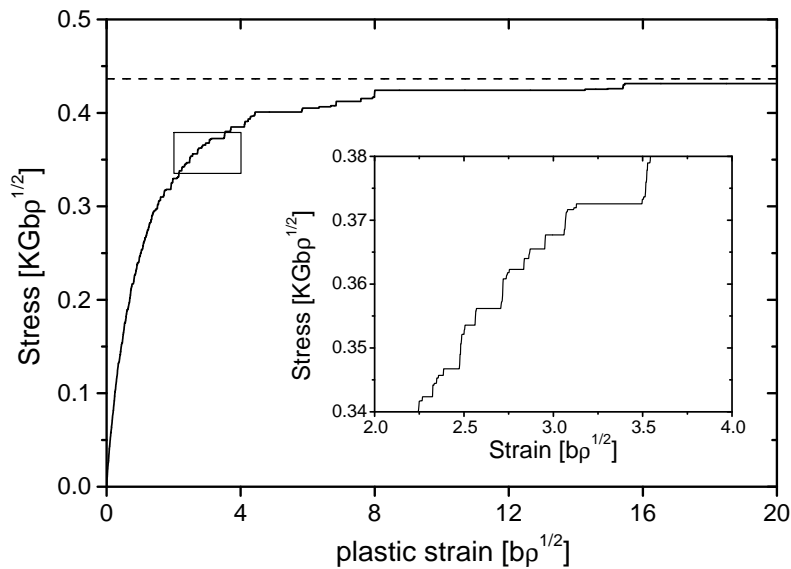,width=8cm,clip=!}}
\caption{Stress-strain curve as obtained from simulation of a
system with 128 $\times$ 128 sites; dashed line: critical stress
$\tau_{\rm c}$; Insert: detail of the same stress-strain
graph.}\label{stressstrain}
\end{figure}
A stress-strain graph obtained for this case is illustrated in
Figure \ref{stressstrain}. The calculation has been done for the
following parameters typical for the onset of deformation of Cu
single crystals: $G = 40 000$ MPa, $\nu = 0.3$, $b=2.5 \times
10^{-10}$m, $\rho = 10^{12}$ m$^{-2}$, and $\xi = 1/\sqrt{\rho} =
10^{-6}$m. With these parameters, $K \approx 1$, while the
parameter $D$ in Eq. (\ref{gradterm}) was chosen as $D=0.1$. We
note that, in the context of the present model, this latter term
mainly serves to break the symmetry which exists in the elastic
kernel given by Eq. (\ref{mf}) between the directions in and out
of the slip plane. The numerical value of $D$ is of little
influence on our results unless very high $D$ values are assumed.

From Figure \ref{stressstrain}, the following observations can be
made: (i) In spite of the fact that the averages of $\delta \tau$
over $\gamma$ at fixed $\r$ (or over $\r$ at fixed $\gamma$) are
zero, a finite stress (yield stress) of $\tau_{\rm c}\sim 0.4$ (in
scaled coordinates) is required to sustain deformation. This is
because at low applied stresses the system becomes pinned in
configurations where the fluctuating stress $\delta \tau$ is
negative in most volume elements, thereby creating a non-zero
average back stress. (ii) The yield stress is approached
asymptotically through a 'microplastic' region where such
metastable configurations are gradually eliminated. (iii) The
increase of plastic strain with increasing stress occurs in
discrete 'slip avalanches' of varying size. These avalanches are
visible as steps on the stress-strain curves, which assume a
staircase-like shape. The intervals between the larger avalanches
divide into avalanches of smaller size, and the characteristic
avalanche size diverges as one approaches the yield stress. In
dimensional units, the scaled stress and strain ranges visible in
Figure \ref{stressstrain} correspond to $0 \le \tau_{\rm ext} \le
5$ MPa and $0 \le \langle \gamma \rangle \le 0.5 \%$,
respectively. The figure can, hence, be thought of as describing
the microplastic region and yielding transition at the onset of
easy-glide deformation of a fcc single crystal. We also note that
the dimensional yield stress obeys Taylor's well-established
scaling relation $\tau_{\rm c} = \alpha G b \sqrt{\rho}$. The
numerical value of $\alpha \approx 0.4$ obtained from the
simulation is within the characteristic range $0.2 < \alpha < 0.5$
observed in experiment.

\begin{figure}[t]
\hspace*{-1.5cm}
\centerline{\epsfig{file=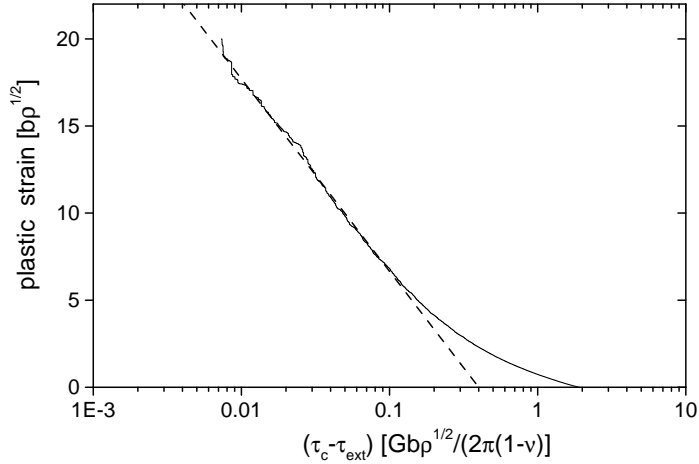,width=7cm,clip=!}}
\caption{Ensemble-averaged stress-strain graph as obtained by
averaging over 60 simulations of systems with 128 $\times$ 128
sites. Strain is plotted against distance from the critical
stress.}\label{stressens}
\end{figure}

We first study the {\em average} behavior of our model. By
averaging the stress-strain graphs over a large number of
simulations, we obtain a (nearly) smooth stress-strain
relationship (Figure \ref{stressens}). The average strain diverges
as the stress approaches the yield stress $\tau_{\rm c}$. A
semi-logarithmic plot of $\gamma$ vs. $\tau_{\rm c}-\tau_{\rm
ext}$ reveals that this divergence is logarithmic in nature.
Accordingly, the stress susceptibility $\chi =
\partial \gamma/\partial \tau_{\rm ext}$ of the plastic strain
diverges according to
\begin{equation}
\chi \propto (\tau_{\rm c}-\tau_{\rm ext})^{-\theta}\;,\;\theta
\approx 1\PP \label{chi}
\end{equation}

\begin{figure}[bt]
\hspace*{-2cm}
\centerline{\epsfig{file=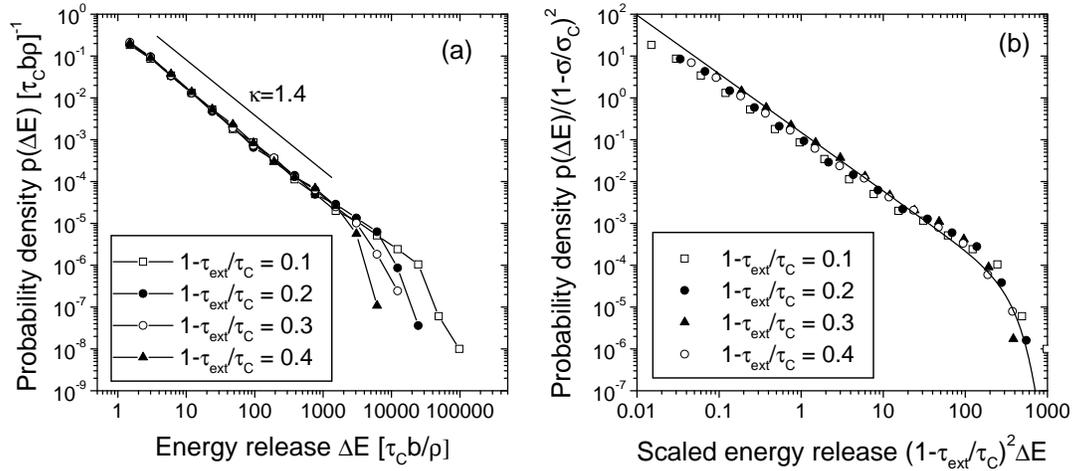,width=10cm,clip=!}}
\caption{Probability distributions of slip avalanche sizes
(probability density $p(\Delta E)$ vs. energy release $\Delta E$)
as obtained from an ensemble of systems of size 256$\times$256.
Left: distributions corresponding to different stresses; Right:
universal distribution obtained by re-scaling $\Delta E \to \Delta
E(1-\tau_{\rm ext}/\tau_{\rm c})^2$; full line: theoretical curve
(Eq. \ref{mfdist}).}\label{avalstat}
\end{figure}

We now investigate in more detail the 'strain bursts' that
manifest themselves as stepwise strain increments in the
individual stress-strain graphs. To this end, we record the strain
increments (if any) which occur after each stress increment
$\Delta \tau_{\rm ext}$. Since the total strain increment due to a
slip avalanche depends on the system size $V_{\rm S}$ (a given
local slip avalanche produces a smaller total strain increment
$\Delta \gamma$ in a larger system),  we characterize the
avalanche sizes in terms of the dissipated energy $\Delta E =
\tau_{\rm ext} \Delta \gamma V_{\rm S}$ which is proportional to
the strain increment but does not depend on system size. We then
determine the probability distribution of avalanche sizes within
different external stress intervals, corresponding to different
distances from the critical stress $\tau_{\rm c}$. Results are
given in Figure \ref{avalstat}.

The distributions exhibit a power-law decay $p(\Delta E) \propto
\Delta E^{-\kappa}$ which is truncated at a characteristic
avalanche size $\Delta E_{\rm c}$. As the stress approaches the
critical stress $\tau_{\rm c}$, this upper limit and hence the
average avalanche size diverge. Distributions obtained at
different applied stresses $\tau_{\rm ext}$ can be collapsed into
a single universal distribution if the energy releases are
re-scaled by a factor $(1-\tau_{\rm ext}/\tau_{\rm c})^{1/\sigma}$
where $\sigma \approx 0.5$ (Figure \ref{avalstat}, right). The
results for the susceptibility and the avalanche statistics are
not independent since the susceptibility is proportional to the
average avalanche size. We therefore expect it to diverge like
$\chi \propto (\tau_{\rm c} - \tau)^{-(2-\kappa)/\sigma}$. With
$\kappa \approx 1.4$ and $\sigma \approx 0.5$ we find $\theta =
(2-\kappa)/\sigma \approx 1$ in agreement with Eq. (\ref{chi}).
The exponents $\kappa$, $\theta$ and $\sigma$ can be understood in
terms of mean-field depinning [see e.g. Fisher (1998) or Zapperi
et al. (1998)], as the theoretical values $\theta = 1, \sigma
=0.5$ and $\kappa = 1.5$ are in good agreement with the present
simulation results. The theoretically predicted avalanche size
distribution for mean-field depinning is
\begin{equation}
p(\Delta E) \propto \Delta E^{-1.5} \exp\left[-\left(\frac{\Delta
E}{\Delta E_{\rm c}}\right)^2\right]\;,\label{mfdist}
\end{equation}
which again compares well with the results of the simulations
(Figure \ref{avalstat}).

\subsection{Influence of hardening}

\begin{figure}[b] \hspace*{-1.5cm}
\centerline{\epsfig{file=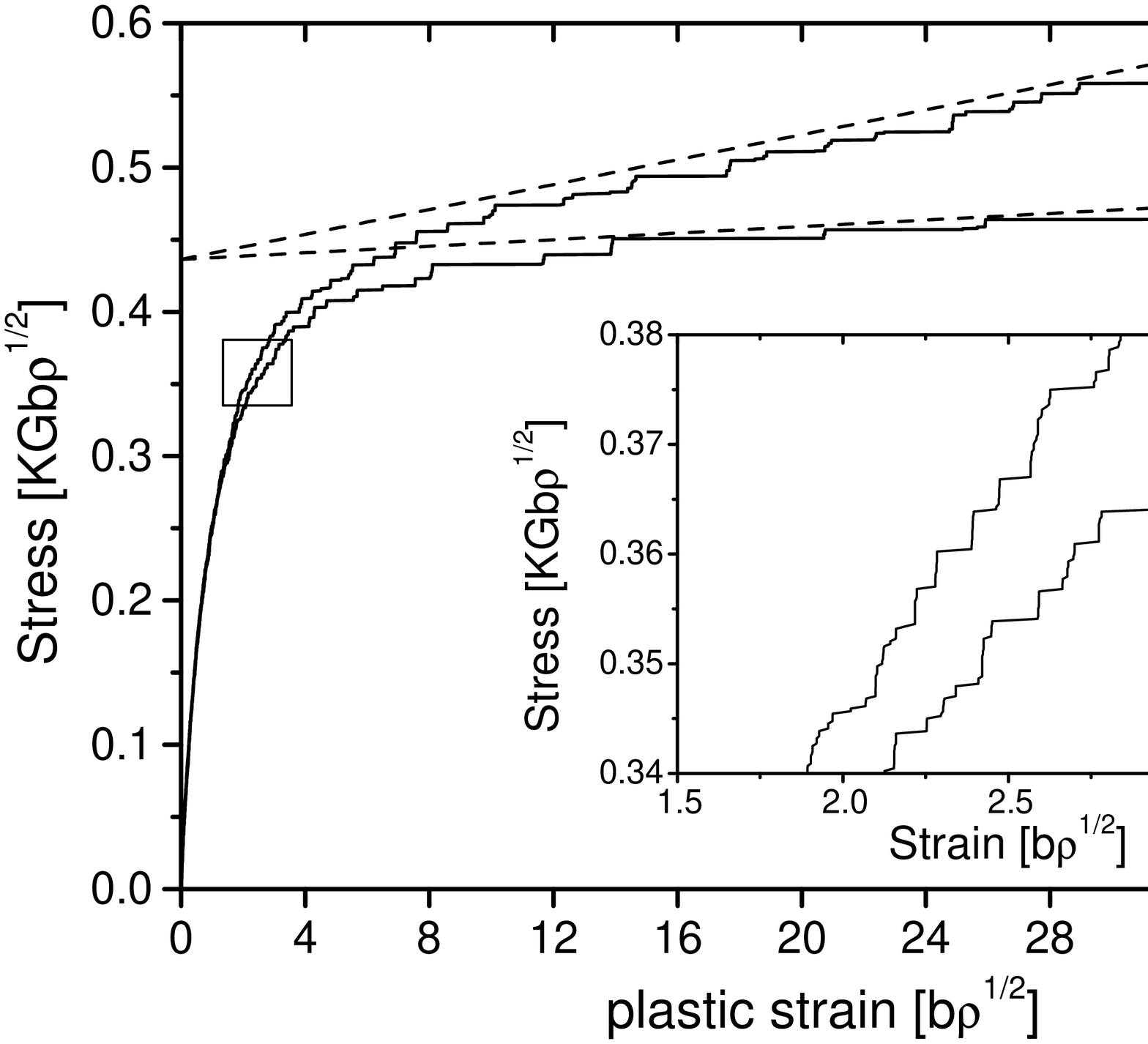,width=8cm,clip=!}}
\caption{Stress-strain curves as in Figure \ref{stressstrain} for
hardening coefficients $\Theta = 0.001$ and $\Theta = 0.004$;
dashed lines: curves $\tau = \tau_{\rm c} + \Theta \gamma$ with
$\tau_{\rm c} = 0.44$.}\label{stressstrainhard}
\end{figure}

To introduce hardening with non-dimensional hardening coefficient
$\Theta$ into our model, we use two approaches: (i) We increase
the amplitude of the internal-stress fluctuations (i.e., the width
of the probability distribution of $\delta \tau$) in proportion
with the local strain by an amount of $\Theta\gamma/\tau_{\rm
c}(\Theta=0)$; (ii) We leave the width of the distribution
unchanged but shift the mean value by an amount $\Theta \gamma$
that is again proportional to the local strain. Both methods
result in stress-strain graphs similar to those depicted in Figure
\ref{stressstrainhard}: The asymptotic behavior changes, as the
horizontal tangent of the stress-strain graph is replaced by a
tangent with slope $\Theta$.
To give again a 'feeling' for the orders of magnitude involved, we
note that the simulated hardening rates correspond, with the
parameters for Cu given above, to dimensional hardening rates of
$40$ and $200 MPa$ which are in the range of what is typically
observed in hardening stages I and II of Cu single crystals.

It is seen from Figure \ref{stressstrainhard} that the influence
of hardening does not change the stepwise morphology of the
stress-strain curves but eliminates the largest strain bursts
which, in a non-hardening system, occur at stresses close to the
critical stress. To quantify this idea, we study the avalanche
statistics (evaluated for stresses above the critical stress of
the non-hardening system) for different hardening rates. Figure
\begin{figure}[bt]
\hspace*{-2cm}
\centerline{\epsfig{file=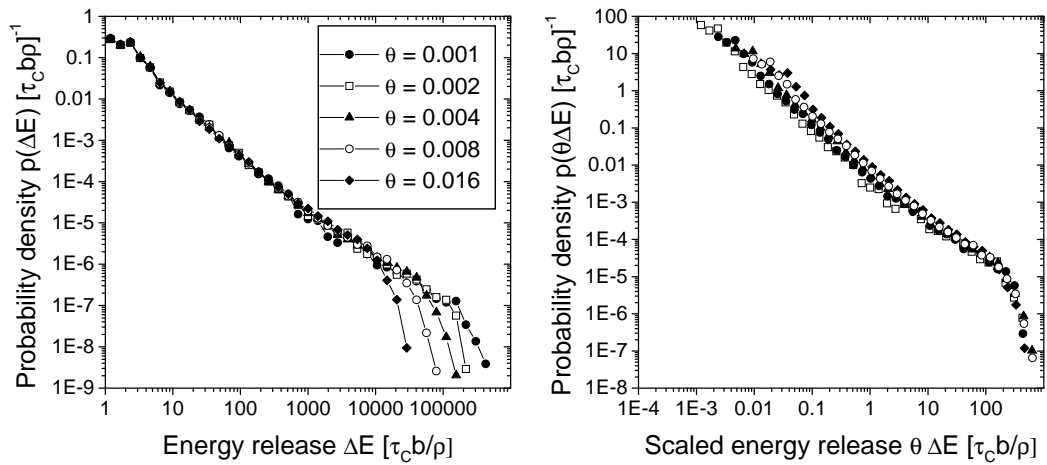,width=10cm,clip=!}}
\caption{Probability distributions of slip avalanche sizes
(probability density $p(\Delta E)$ vs. energy release $\Delta E$)
as obtained from an ensemble of systems of size 256$\times$256.
Left: distributions corresponding to different hardening rates;
Right: universal distribution obtained by re-scaling $\Delta E \to
\Theta \Delta E$}\label{avalhard}
\end{figure}
\ref{avalhard} shows that the hardening does not change the
power-law regime of the burst size distribution but decreases the
cut-off value, which turns out to be inversely proportional to the
hardening rate.

This can be easily understood from a scaling argument. The burst
energy is proportional to the strain produced in a burst, and with
an avalanche exponent of $\kappa \approx 1.5$ it follows that the
average strain $\Delta \gamma$ follows the relation $\langle
\Delta \gamma \rangle \propto \langle \Delta E \rangle \propto
\Delta E_{\rm max}^{1/2}$. Due to hardening, each burst raises the
critical stress by a small amount $\Delta \tau_{\rm c} \propto
\Theta \langle \Delta \gamma \rangle$, which implies that the
external stress (which cannot instantaneously follow) lags behind
the current critical stress by a similar amount. Finally, the
maximum burst size follows the scaling relationship $\Delta E_{\rm
max}\propto \Delta \tau^{-2}$. Combining these relations, it
follows directly that $\Delta E_{\rm max} \propto 1/\Theta$.

\subsection{Slip pattern and surface morphology}

As shown in the previous sections, mean-field depinning provides a
framework for understanding the size distribution of slip
avalanches and the behavior of the stress-strain curves as one
crosses the microplastic region and approaches the yield stress,
as well as for assessing the influence of strain hardening.
However, mean-field considerations can in principle not account
for the observed anisotropic spatial distribution of slip, the
formation of slip lines, and the development of a self-affine
surface morphology. However, our simulations indicate that, owing
to the particular properties of the elastic kernel, such features
naturally emerge from the dynamics of our model.

\begin{figure}[t]
\hspace*{-2cm}
\centerline{\epsfig{file=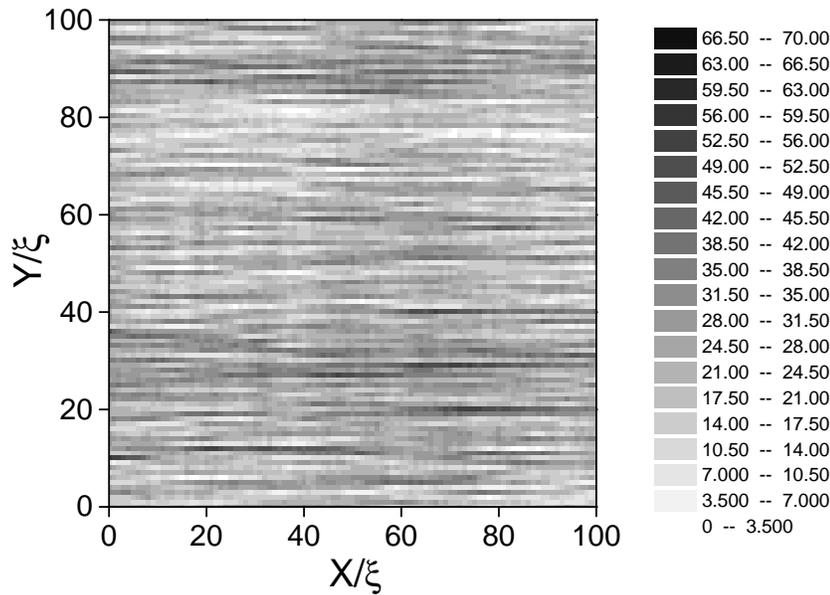,width=8cm,clip=!}}
\caption{Strain pattern obtained after simulation of a system of
size $256 \times 256$ to an average strain of $20 b \sqrt{\rho}$
(slip direction from left to right); parameters as in Figure
\ref{stressstrain}; greyscale: local strain in units of
$b\sqrt{\rho}$} \label{strainpat}
\end{figure}

Figure \ref{strainpat} shows a grey-scale representation of the
slip pattern $\gamma(x,y)$ that emerges in a typical simulation.
One observes a 'slip-line'-pattern with pronounced correlations
along the direction of slip ($x$ direction in Figure
\ref{strainpat}). Such stratified slip patterns are commonly
observed in deformation of single-slip oriented single crystals.
They are considered to be a natural consequence of the directed
glide of lattice dislocations along the slip planes. The present
model produces 'slip-line' patterns without explicitly accounting
for the motion of lattice dislocations. This is can be understood
by noting that according to Eq. (\ref{stress2dfour}))
heterogeneities of slip in the direction of the slip plane normal
(the $y$ direction) do not give rise to any long-range stresses.
Heterogeneities in the slip direction (the $x$ direction), on the
other hand, are penalized by the gradient-dependent stress
contribution in Eq. (\ref{depin1}) which tends to homogenize
deformation in the slip direction. Because of these properties of
the interactions, the 'manifold' $\gamma(x,y)$ is prone to fall
apart in the $y$ direction, leading to the striated deformation
patterns seen in Figure \ref{strainpat}.

\begin{figure}[t]
\hspace*{-2cm}
\centerline{\epsfig{file=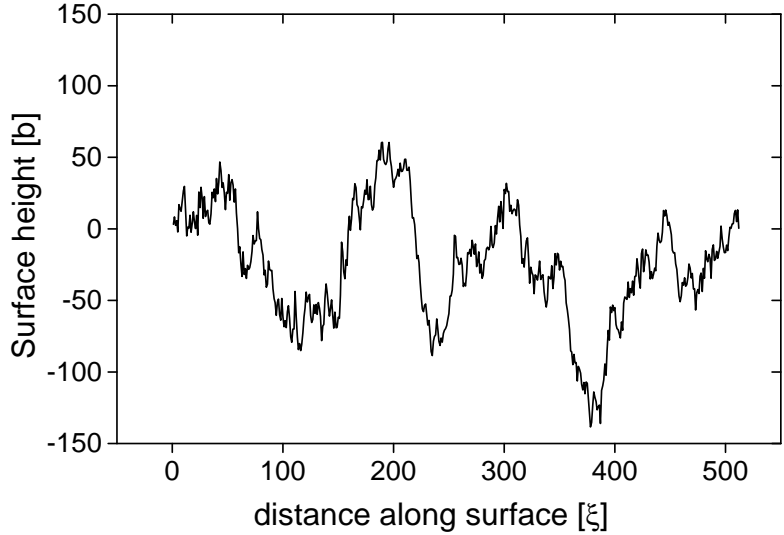,width=8cm,clip=!}}
\caption{Surface profile obtained from a system of size $512
\times 512$ after simulation to an average strain of
$\bar{\gamma}=20 b \sqrt{\rho}$.} \label{profilesim}
\end{figure}

We now evaluate the surface profiles which correspond to the
simulated slip patterns. We consider a surface running along the
plane $x=0$. The displacement $u$ in the direction normal to the
surface fulfils the relationship
\begin{equation}
\frac{\partial u}{\partial y} = \gamma(x=0,y)\;.
\end{equation}
This can be understood as the sum of a displacement due to the
average strain $\langle \gamma \rangle$, which leads to a rigid
rotation of the surface, plus a random displacement due to the
strain fluctuation $\delta \gamma(x=0,y) = \gamma(x=0,y) - \langle
\gamma \rangle$. The surface profile may be obtained by direct
integration of the strain fluctuation,
\begin{equation}
h(y) = \int_0^y [\gamma(x=0,y')-\langle \gamma \rangle] \d y'\;.
\end{equation}
A surface profile obtained in this manner is shown in Figure
\ref{profilesim}.

\begin{figure}[t]
\hspace*{-2cm}
\centerline{\epsfig{file=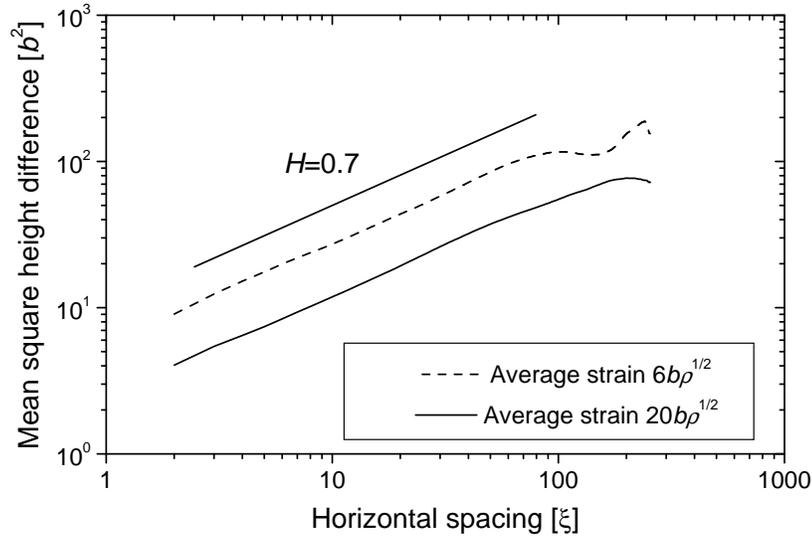,width=8cm,clip=!}}
\caption{Mean square height difference vs. horizontal distance for
surface profiles obtained from a system of size $512 \times 512$
after simulation to average strains of $6 b \sqrt{\rho}$ and $20 b
\sqrt{\rho}$; each graph has been averaged over 10 simulated
profiles.} \label{roughsim}
\end{figure}

To assess simulated profiles in terms of their roughness exponent,
we plot the mean square height difference $\langle |u(x)-u(x+l)|
\rangle$ between two points along the profile as a function of
their separation $l$. For a self-affine profile we expect $\langle
|u(x)-u(x+l)| \rangle \propto l^H$ where $H$ is the Hurst
exponent. Figure \ref{roughsim} demonstrates that our simulated
profiles can be characterized by a Hurst exponent $H \approx 0.7$.
Increasing the total strain leads to an increase in the absolute
magnitude of surface height variations but does not change the
value of $H$.
Both the numerical value of the Hurst exponent and its strain
independence compare well with the experimental observations
quoted in the introduction.

\section{Conclusions}

Our simulations demonstrate that it is possible to capture
essential aspects of slip avalanches in terms of a simple
continuum model. By combining a fluctuating local stress-strain
relationship with a strain-gradient dependent stress contribution
and long-range stresses mediated by the elastic Green's function,
the model accounts both for the observed spatial heterogeneities
of plastic deformation (slip lines, self-affine surface roughness)
and the emergence of 'bursts' of plastic activity with a power-law
size distribution.

The present model has a fairly generic formal structure, and many
of our qualitative results do not depend critically on the
particular physical assumptions made. For instance, it is not
crucial to relate the fluctuating flow stress to the stress
fluctuations associated with a homogeneous distribution of
dislocations as done in Section 2.3. Concepts of randomness and
disorder also apply to flow stress fluctuations arising from other
types of random microstructure such as irregular patterns of
dislocation cells, or from the statistical activation and
exhaustion/blocking of dislocation sources. However, there are
still several crucial questions that need to be studied: (i) Is it
possible to formulate a dislocation-based model which exhibits the
same behavior as the present phenomenological continuum model?
(ii) How does the behavior of the model change if we allow for
slip on multiple slip systems, such that the tensorial nature of
the plastic strain becomes important? (iii) What is the influence
of surface boundary conditions on the slip patterns -- are the
simulated 'surface profiles' truly equivalent to those emerging on
a free surface, where the stresses are by necessity different from
the bulk?

In general terms, or results indicate that fluctuation phenomena
in plastic flow of crystalline solids can be envisaged as critical
phenomena in driven non-equilibrium systems. The theory of elastic
interface depinning provides a theoretical framework which covers
a wide range of phenomena including avalanche dynamics in the
motion of magnetic domain walls (Barkhausen noise, Zapperi et al.
1998), the dynamics of earthquakes or the motion of cracks
(Fisher, 1998). The present model adopts this framework to
problems of plastic flow. An analogous effort has been made in
relation with the plasticity of amorphous materials by Baret et
al. (2002). Peculiarities arise in the present case from the
particular deformation geometry: In a crystalline solid deforming
in single slip, where deformation is restricted to simple shear
occurring along a single set of planes, the elastic kernel which
mediates long-range elastic interactions is not positively
definite. The consequences of this observation as well as the
reasons for the emergence of a power-law decay of correlations in
the strain pattern (which has, to the knowledge of the authors,
not been observed in other depinning-type models) remain to be
assessed in future work.

\section*{Acknowledgements}
Financial support by the European Commission under RTN/SizeDepEn
HPRN-CT 2002-00198 and of EPSRC under Grant No. GR/S20406/01 is
gratefully acknowledged. We also thank Mikko Alava and Stefano
Zapperi for many stimulating discussions on problems of interface
depinning.

\section*{References}

Baret, J.-C., Vandembroucq, D., Roux, S., 2002. An extremal model
for amorphous media plasticity. Phys. Rev. Lett. 89, 195506.

Chen,K., Bak, P., Obukhov, S.P., 1991. Self-organized criticality in
a crack-propagation model of earthquakes. Phys. Rev. A 43, 625-630.

Fisher, D.S., 1998. Collective transport in random media: From
superconductors to earthquakes. Phys. Rep. 301, 113-150.

Groma, I., Bako, B., 1998. Probability distribution of internal
stresses in parallel straight dislocation systems. Phys. Rev. B
58, 2969-2974.

Groma, I., Csikor, F., and Zaiser, M., 2003. Spatial correlations
and higher-order gradient terms in a continuum description of
dislocation dynamics. Acta Mater. 51, 1271-1281.

Middleton, A.A., 1992. Asymptotic uniqueness of the sliding state
for charge-density waves. Phys. Rev. Lett. 68, 670-673.

Miguel, M.C., Vespignani, A., Zapperi, S., Weiss, J., Grasso,
J.-R., 2001. Intermittent dislocation flow in viscoplastic
deformation. Nature 410, 667-671.

Neuh\"auser, H., 1984. Slip-line formation and collective
dislocation motion. In: F.R.N. Nabarro (Ed.), Dislocations in
Solids, Vol. 4, North-Holland, Amsterdam, pp. 319-440.

Weiss, J., Grasso, J.-R., 1997. Acoustic emission in Single
Crystals of Ice. J. Phys. Chem. B 101, 6113-6117.

Weiss, J., Marsan, D., 2003. Three-dimensional mapping of
dislocation avalanches: Clustering and space/time coupling.
Science 299, 89-92.

Uchic, M.D., Dimiduk, D.M., Florando, J.N., and Nix, W.D., 2004.
Sample Dimensions Influence Crystal Strength and Plasticity.
Science 305, 986-989.

Zaiser, M., 1998. A generalized composite approach to the flow
stress and strain hardening of crystals containing heterogeneous
dislocation distributions. Materials Science and Engineering A
249, 145-151.

Zaiser, M., Miguel, M.-C, Groma, I., 2001. Statistical dynamics of
dislocation systems: The influence of dislocation-dislocation
correlations. Phys. Rev. B 64, 224102.

Zaiser, M., Seeger, A., 2002. Long-range internal stresses,
dislocation patterning and work hardening in crystal plasticity.
In: F.R.N. Nabarro and M.S. Duesbery (Eds.), Dislocations in Solids,
Vol. 11, North-Holland, Amsterdam, pp. 1-100.

Zaiser, M., Madani F., Koutsos, V., Aifantis E.C., 2004.
Self-affine surface morphology of plastically deformed metals.
Phys. Rev. Letters 93, 195507.

Zaiser, M., Aifantis, E.C., 2005. On the theory of gradient
plasticity III: Random effects and slip avalanches, Int. Journal
of  Plasticity (in press).

Zapperi, S., Cizeau, P., Durin, G., Stanley, H.E., 1998. Dynamics
of a ferromagnetic domain wall - Avalanches, depinning and the
Barkhausen effect. Phys. Rev. B 58, 6353-6366.

\end{document}